\documentclass[showpacs,superscriptaddress,prb,preprint]{revtex4-1}
\usepackage{natbib}
\usepackage{graphicx}
\usepackage{amssymb}
\usepackage{amsmath}
\usepackage{bm}

\begin{document}

\title{Twisted phase of the orbital-dominant ferromagnet SmN
  in a GdN/SmN heterostructure}

\author{J.~F.~McNulty}
\author{E.~-M.~Anton}  
\author{B.~J.~Ruck}
\author{F.~Natali} 
\author{H.~Warring}

\affiliation{The MacDiarmid Institute for Advanced Materials and
  Nanotechnology, School of Chemical and Physical Sciences, Victoria
  University of Wellington, P.O. Box 600, Wellington, New Zealand}

\author{F.~Wilhelm}
\author{A.~Rogalev}
\author{M.~Medeiros Soares}
\author{N.~B.~Brookes}

\affiliation{ESRF-The European Synchrotron, CS40220, F-38043 Grenoble
  Cedex 9, France}

\author{H.~J.~Trodahl} 

\affiliation{The MacDiarmid Institute for Advanced Materials and
  Nanotechnology, School of Chemical and Physical Sciences, Victoria
  University of Wellington, P.O. Box 600, Wellington, New Zealand}

\begin{abstract}
  The strong spin-orbit interaction in the rare-earth
  elements ensures that even within a ferromagnetic state
  there is a substantial orbital contribution to the
  ferromagnetic moment, in contrast to more familiar
  transition metal systems, where the orbital moment is
  usually quenched. The orbital-dominant magnetization that
  is then possible within rare-earth systems facilitates the
  fabrication of entirely new magnetic heterostructures, and
  here we report a study of a particularly striking example
  comprising interfaces between GdN and SmN. Our
  investigation reveals a twisted magnetization arising from
  the large spin-only magnetic moment in GdN and the nearly
  zero, but orbital-dominant, moment of SmN. The unusual
  twisted phase is driven by (i) the similar ferromagnetic
  Gd-Gd, Sm-Sm and Gd-Sm exchange interactions, (ii) a SmN
  Zeeman interaction 200 times weaker than that of GdN, and
  (iii) the orbital-dominant SmN magnetic moment.  The
  element specificity of X-ray magnetic circular dichroism
  (XMCD) is used in seperate modes probing both bulk and
  surface regions, revealing the depth profile of the
  twisting magnetization.

\end{abstract}


\date{\today}
\maketitle
\section{Introduction}
An inhomogeneous, twisted magnetic ordering commonly occurs
near interfaces between ferromagnetic materials, due to
competing interactions which favor opposing alignments of
the magnetization. These phases are types of engineered
domain walls, and thus have important implications for
spintronics applications, where current-driven domain wall
motion is an active area of
research.\cite{Parkin_Science_2008,Thiaville_EPL_2012,
  Emori_Nature_2013,Khvalkovskiy_PRB_2013} So far, twisted
phases are known to manifest in diverse magnetic systems,
\cite{Zocher_TFS_1933,
  Thiaville_JMMM_1992,Camley_PRB_1987,Goto_JAP_1965,
  Berkowitz_JMMM_1999, Bogdanov_PRB_2003,
  Bogdanov_PRL_2001,Rossler_Nature_2006} however these all
fall under the conventional spin-dominant paradigm of
magnetism where the orbital moment plays no significant
role. Competing interactions in the presence of a dominant
orbital moment have so far remained unexplored, yet the
opportunity now exists within the rare-earth nitride (REN)
series, where orbital-dominant magnetism is possible due to
strong spin-orbit coupling of the $4f$ electrons. Forming a
series of mostly intrinsic ferromagnetic semiconductors,
\cite{Natali_PMS_2013,Leuenberger_PRB_2005,Granville_PRB_2006,
  Preston_PRB_2007,Meyer_JMMM_2010,Azeem_JAP_2013,Binh_PRL_2013}
the RENs are already integrated within spintronic devices,
\cite{Senapati_Nature_2011,Muduli_PRB_2014} and thus provide
a novel system for studying competing interactions.

The rare-earth elements, comprising the series across which
the 4$f$ shell is filled, have been of interest for nearly a
century. They are most commonly found in the trivalent state
in a wide range of compounds, including the RENs. The 4$f$
shell, with $l = 3$, comprises seven distinct orbital
states, $-3 \leqslant m_l \leqslant 3$, and with the spin
degeneracy a total of 14 single-electron states.  Gd$^{3+}$
has a half filled shell, for which Hund's rules state that
the seven electrons fill all of the orbital states with
spin-up electrons; $L = 0$ and $S = J = 7/2$. It thus has a
purely spin moment of 7~$\mu_B$. The indirect exchange
interaction aligns the spins below a Curie temperature of
about 50~K, rising to 70~K under heavy donor
doping,\cite{Natali_PRB_2013} but the spherical symmetry of
the $L = 0$ shell interacts very weakly with the crystalline
environment, leading to a coercive field as small as
100~Oe.\cite{Ludbrook_JAP_2009} Sm provides an enormous
contrast. In the Sm$^{3+}$ ion there are 5 electrons in the
4$f$ shell, again with full spin alignment ($S = 5/2$) in
the Hund's rule ground state, and with an orbital angular
momentum $L = 5$, opposing the spin. The simple Hund's rule
result is then that the magnetic moment of the 4$f$ shell is
$\mu = \mu_B\langle L_z+2S_z\rangle = 0$.  As usual the
spin-orbit interaction prevents the multi-electron state
from adopting fixed $m_s$ and $m_l$, and the free Sm$^{3+}$
ion has the paramagnetic moment defined by the Land\'{e}
$g$-factor via $g\sqrt{J(J+1)}\mu_B = 0.84$~$\mu_B$ per
ion. In SmN that is reduced by the crystal field to
0.41~$\mu_B$, as reported by Meyer \emph{et
  al.}\cite{Meyer_PRB_2008} However, the net moment in the
ferromagnetic state below 27~K is only 0.035~$\mu_B$ per
Sm$^{3+}$, and is directed antiparallel to the spin moments
that are aligned by inter-ion
exchange.\cite{Meyer_PRB_2008,Anton_PRB_2013} The orbital
moment is then parallel to the net moment, and SmN is
properly viewed as an orbital-dominant ferromagnet. The
coercive field in SmN is enhanced to over 6~T by the
non-spherical $L = 5$ orbital wave function and the very
weak Zeeman interaction associated with the small magnetic
moment.

\begin{figure}
\center \includegraphics[width=0.75\columnwidth]{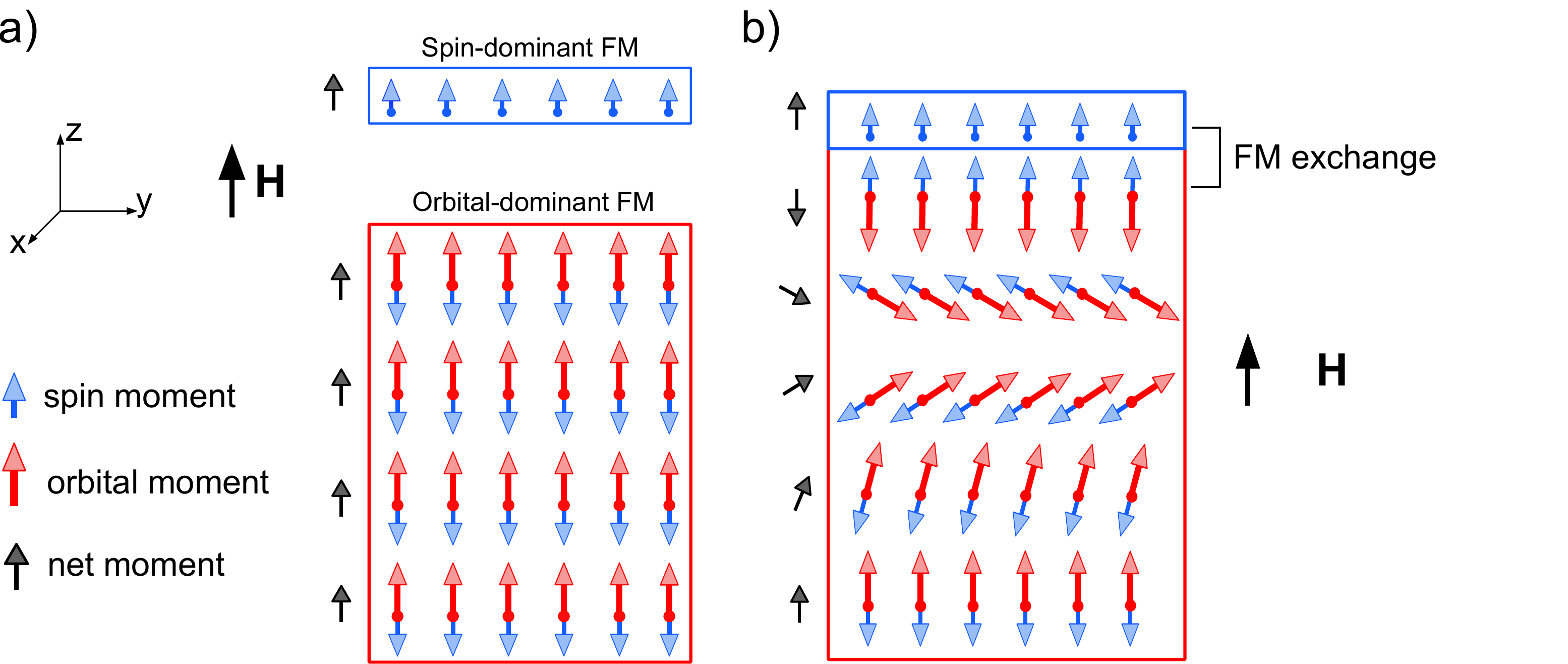}
\caption{\label{fig1}(color online) (a) A sketch of a single atomic
  layer of a spin-dominant ferromagnet (e.g. GdN) and a cross section
  of multiple atomic planes of an orbital-dominant (SmN)
  ferromagnet. (b) Cross-section of an interface between spin and
  orbital-dominant ferromagnets. A twisted phase develeps in the
  orbital-dominant magnet due to exchange-Zeeman competition which
  occurs if the spin-dominant layer remains fixed due to its large
  Zeeman coupling.}
\end{figure}

Here we exploit the contrasting properties of GdN and SmN in SmN/GdN
thin film heterostructures, and observe a twisted phase arising from a
novel competition between spin and orbital magnetism. The
spin-dominant GdN is fixed parallel to an external magnetic field, and
its much larger Zeeman interaction ensures that it provides a rigid
layer which pins the SmN spin at the SmN-GdN interface. The pinning of
the SmN, with its 200-fold weaker Zeeman coupling, takes place through
ferromagnetic exchange coupling with the GdN, resulting in a SmN
spin-moment parallel to that of the GdN, while the orbital-moment is
antiparallel. This interface pinning is opposed by the
orbital-dominant Zeeman alignment of the bulk SmN, which tends to
align the SmN magnetization in the opposite sense, and thus drives the
rotation of the magnetization across the SmN layer. Figure \ref{fig1}
sketches the effects of exchange coupling between spin and orbital
dominant ferromagnets.

It is important to note that the GdN/SmN system is
fundamentally different from the conventional spin-dominant
ferromagnetic systems displaying twisted phases. The most
common exchange spring systems, composed of hard and soft
ferromagnetic layers, are first magnetized in one direction,
and when the field is reversed the hard material remains
fixed while an exchange spiral is formed in the soft
material.\cite{Goto_JAP_1965,Fullerton_PRB_1998} In another
manifestation, metallic Gd/Fe systems displaying twisted
phases rely on antiferromagnetic coupling between spins at
the
interface.\cite{Camley_PRB_1987,Camley_PRB_1988,Dufour_PRB_1993,Hahn_PRB_1995,
  Kravtsov_PRB_2009,Koizumi_PRB_2000} With the SmN/GdN
system, however, the interlayer Sm-Gd exchange is
ferromagnetic, and the usual hard/soft contrast is of no
interest; indeed the fixed layer (GdN) has a coercive field
three orders of magnitude smaller than SmN. It is the much
stronger Zeeman interaction in GdN than in SmN that
effectivel locks the GdN magnetization.  Furthermore, the
spin-dominant, metallic systems lack the novel combination of
electronic and magnetic properties of SmN and GdN, which
allow the facility of controlling the concentration and sign
of charge carriers without disturbing the ferromagnetic
ordered state, and band structure results also show electron
and hole channels of majority spin.\cite{Larson_PRB_2007}

In our investigation of the interface exchange coupling in
GdN/SmN multilayers we have used the element selectivity of
X-ray magnetic circular dichroism (XMCD) at the Sm L$_{2,3}$
and M$_{4,5}$ edges. We first demonstrate that the SmN is
ferromagnetically exchange coupled to GdN through
investigation of a SmN/GdN superlattice. We then demonstrate
that a twisted, or rotating, magnetization develops in
ultrathin SmN films coupled to GdN due to interface pinning
in the SmN, short-range interionic rare-earth exchange, and
the extremely weak Zeeman coupling of SmN. The observed
depth dependence of the magnetization is fully consistent
with an analytical model based on these competing
interactions.

\section{Experimental Details}
The attenuation lengths of hard L-edge and soft M-edge
X-rays dictated that quite different structures were used
for the two investigations. At the L-edge the full thickness
of a superlattice of 12$\times$(1.5 nm SmN/9 nm GdN) was
probed through a 100 nm passivating AlN cap. For the much
more surface sensitive M-edge we investigated two
samples. The first was a bilayer of 100 nm GdN/ 5.5 nm SmN,
and the second, a trilayer of 100 nm GdN/ 6 nm LaN/ 5.5 nm
SmN. The non-magnetic LaN layer between the GdN and SmN was
included to block the Gd-Sm exchange interaction in the
trilayer. Both of the M-edge samples were passivated with 25
nm of GaN to prevent sample oxidation.

Samples were grown in a Thermionics ultra-high vacuum system
with a base pressure of $1\times 10^{-8}$ Torr. High purity
Gd metal was evaporated at a rate of 0.2 \AA/s with a N$_2$
partial pressure of $4.5\times 10^{-4}$ Torr. Sm
metal was evaporated at a rate of 0.3 \AA/s under the same
N$_2$ pressure. The superlattice was grown on an MgO(111)
substrate, while the bi- and trilayers were grown on c-plane
Al$_2$O$_3$ substrates. All the substrates were outgassed
for 1 hour at 700~$^{\circ}$C, and heated to 600~$^{\circ}$C
during growth. The GaN and AlN capping layers were grown at
room temperature with the metal evaporated at a rate of 0.1
\AA /s with an ion source activating the N$_2$. Thicknesses
were determined via quartz crystal balances calibrated for
SmN, GdN, AlN, and GaN via scanning electron microscope and
Rutherford backscattering measurements. The SmN/GdN
superlattice was characterized \emph{ex situ} by XRD, and
showed the lattice constant of GdN; as expected the in-plane
lattice constant was dominated by the thicker GdN layers in
all cases.

Magnetization measurements were carried out via a Quantum
Design SQUID with the field oriented in-plane. Because the
much larger magnetic moment of GdN drowns out the signal
from SmN, SQUID measurements probe only the GdN
magnetization. Curie-Weiss fits to the inverse
susceptibility yielded paramagnetic Curie temperatures of 69
K, 68 K and 66 K for the superlattice, trilayer, and
bilayer, respectively. Hysteresis loops measured at 5 K
saturated at 7 $\mu_B$ per Gd$^{3+}$ ion. The superlattice
and bilayer displayed a coercive field of 120 Oe at 5 K
while the trilayer had a coercive field of 90 Oe, all within
the range reported for polycrystalline GdN films.
\cite{Ludbrook_JAP_2009, Natali_JCG_2010}

XMCD measurements were performed at temperatures down to
15~K and fields up to 6~T at the Sm and Gd L$_{2,3}$ edges
on beam line ID12 at the European Synchrotron Radiation
Facility (ESRF) in Grenoble. M$_{4,5}$ edge XMCD was
measured at the soft X-ray line ID08 of the ESRF, at
temperatures to 10 K and in fields up to 4 T. Measurements
at the M-edge were necessarily performed only at normal
incidence to limit attenuation by a passivating cap. For all
of the L-edge XMCD measurements the field and incident beam
were directed at 10$^{\circ}$ from grazing incidence, in
which geometry the very large shape anisotropy ($4\pi
\mathcal{M} \sim 2$ T, were $\mathcal{M}$ is the
magnetization) of GdN ensured that the magnetization lay in
the plane of the film. At both edges the applied magnetic
field was along the X-ray propagation direction.

The XMCD spectra were obtained by taking the difference of
two XAS spectra with the X-ray helicity reversed while the
magnetic field was held fixed. This corresponds to the
difference between antiparallel and parallel alignments of
the helicity and magnetization. XAS spectra have been
normalized to the incident photon intensity. XMCD spectra
were normalized to the XAS white line intensity at the
M-edge and to the edge-jump at the L-edge. XAS and XMCD
spectra for Sm and Gd L-edges appear in the Supplementary
Material (Figure S1), along with Sm and Gd M$_{4,5}$-edge
XAS (Figures S2 \& S3).

XMCD at the Sm L$_2$ edge is the signal of choice for
following magnetic hysteresis, for it is stronger than the
L$_3$ edge signal. In the superlattice that feature was
obscured by magnetic EXAFS (extended X-ray absorption fine
structure) from the Gd L$_3$ edge, necessitating the use of
the Sm L$_3$ XMCD in the superlattice. There was a similar
interference in the soft-X-ray measurements, where the
capping-layer Ga L$_{2,3}$ edge introduced a large and
variable background in the Gd M$_{4,5}$-edge XAS.

Our investigation relies on the use of two common schemes
for measuring the X-ray absorption and XMCD spectra, based
on the emission of fluorescence (total fluorescence yield,
TFY) or electrons (TEY). Below we exploit the differing
probing depths of these two schemes in our soft-X-ray
M$_{4,5}$ edges, where TFY probes the full 5.5 nm of the SmN
layers while TEY data probe a depth of $\sim 2$
nm. Saturation effects distorted the TFY mode at the Sm
M$_{4,5}$ edges, but nonetheless provide relative
comparisons between different samples.

\section{L-edge XMCD Results}

\begin{figure*}
\center \includegraphics[width=\textwidth]{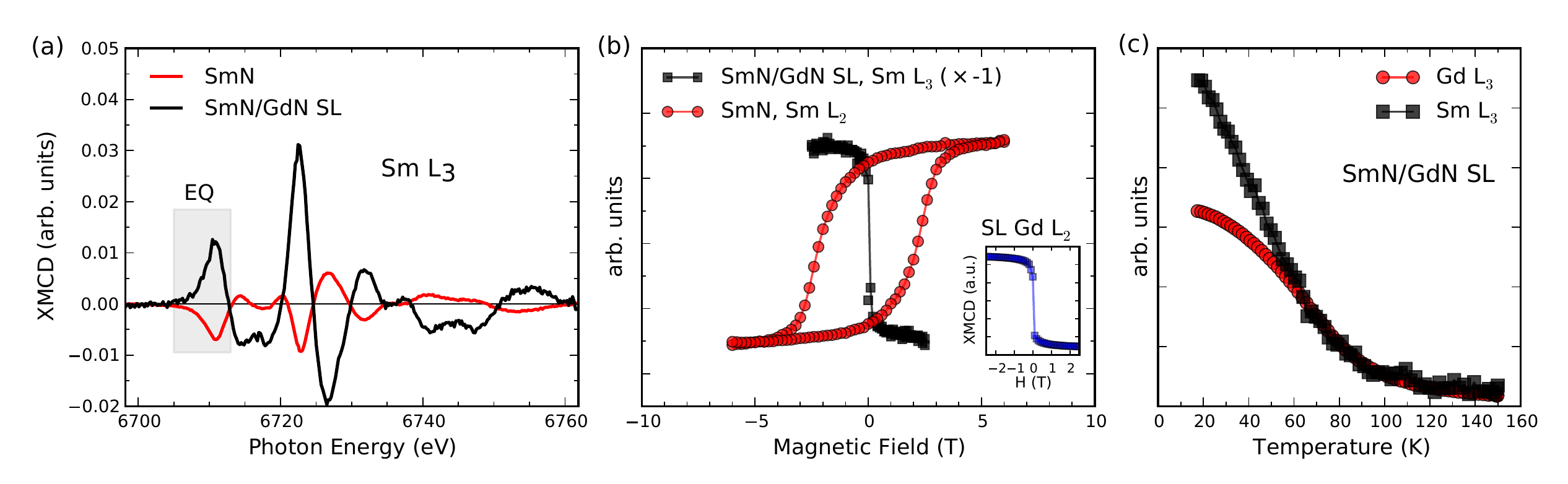}
\caption{\label{fig2}(color online) (a) XMCD at the Sm L$_3$
  edge in a SmN/GdN superlattice (black) and a homogeneous
  SmN film (red) taken at 15~ K, and in
  grazing-incidence. The signal above 6715 eV is
  predominantly due to electric dipole transitions into the
  5$d$ shell, and below that the signal is due to electric
  quadrupole transitions (EQ) into the empty 4$f$
  orbitals. (b) XMCD-derived hysteresis taken at 15 K and
  measured at the Sm L$_2$ edge for the superlattice
  (squares) and at the Sm L$_3$-edge for the bulk SmN film
  (circles) The superlattice spectra were scaled by -1. The
  inset shows the Gd L$_2$ edge hysteresis taken at
  15~K. (c) Temperature dependence of Gd L$_3$ and Sm L$_3$
  peaks in a field of 2.5~T for the SmN/GdN superlattice.}
\end{figure*}

We first discuss the hard X-ray results; Figure
\ref{fig2}(a) shows XMCD data from the superlattice at the
Sm L$_3$ edge, compared to the Sm L$_3$ in homogeneous
SmN. The XMCD spectra from these samples are taken from
Ref. \onlinecite{Anton_PRB_2013} (see also the Supplementary
Material) These spectra primarily show the dipole
transitions from 2$p$ to empty 5$d$ orbitals, with weaker
quadrupolar excitations to the 4$f$ shell, and thus signal
the strength and sign of the spin and orbital alignments of
the 5$d$, and less quantifiably, the 4$f$ shells. The 5$d$
states participate in the ordering through 4$f$-5$d$
exchange, though the exchange mechanism between 5$d$ states
is not well understood.\cite{Duan_JPCM_2007} The XMCD sign
reversal shown in Fig.~\ref{fig2}(a) between homogeneous SmN
and thin SmN layers embedded in GdN immediately indicates
that Sm-Gd interface exchange determines the Sm spin
alignment, dominating the weak Zeeman interaction that
aligns the net, orbital-dominated, moment in homogeneous
SmN.

The hysteresis displayed in Fig.~\ref{fig2}(b) compares the
hysteresis between homogeneous SmN and SmN in the
superlattice, further demonstrating that the SmN coercive
field in the superlattice is reduced to $\sim 0.01$ T,
emphasizing that the SmN magnetization is firmly coupled to
the GdN by the exchange interaction across the GdN/SmN
interfaces. We note that the hysteresis in the homogeneous
SmN film was measured using the L$_2$ edge, the signal of
choice for its substantially larger XMCD signal, but the
masking of that signal by Gd magnetic EXAFS (extended X-ray
absorption fine structure) dictated the use of the weaker
L$_3$ edge in the superlattice. There is an intrinsic sign
difference between the most prominant XMCD features at the
Sm L$_2$ and L$_3$ edges,\cite{Anton_PRB_2013} so we have
scaled the L$_3$ derived hysteresis by -1 in
Fig. \ref{fig2}(b) in order to indicate the antiparallel
spin/orbit alignment between samples, which is clear from
the direct L$_3$ edge comparison in Fig. \ref{fig2}(a).

The temperature dependencies of the GdN L$_3$ and SmN L$_3$
XMCD from the superlattice are compared in
Fig.~\ref{fig2}(c), showing Sm alignment following Gd well
above the 27 K SmN Curie temperature. Clearly the Sm moments
in interface-adjacent ions are again aligned across the
interface. At lower temperatures the Sm moment continues to
rise faster than does the rapidly saturating GdN, as the Sm
ions deeper in the SmN layer align by the Sm-Sm exchange
interaction.

\section{M-edge XMCD Results}

\begin{figure*}
\center \includegraphics[width=\textwidth]{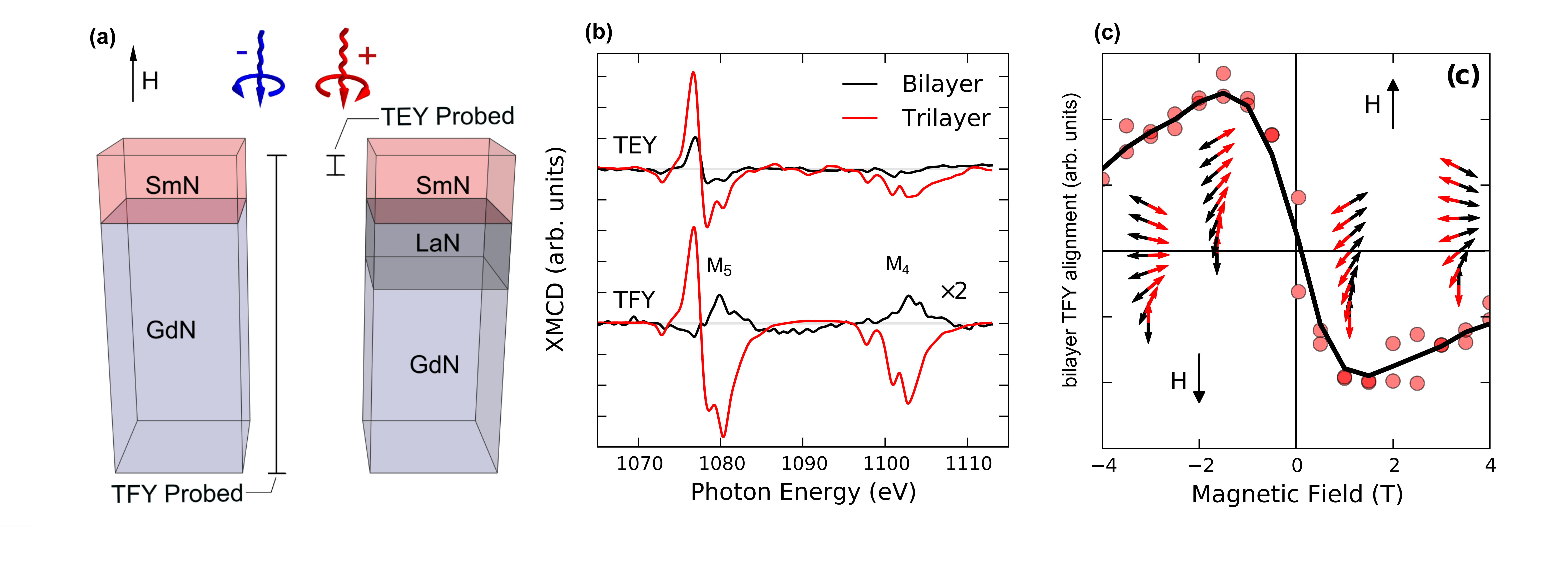}
\caption{\label{fig3}(color online) (a) Bilayer and trilayer
  field-normal XMCD geometry at the M$_{4,5}$ edge, with the
  approximate probing depths of the XMCD signal in the TEY
  and TFY detection modes sketched. (b) The M$_{4,5}$-edge
  XMCD spectra for the bilayer (black) and trilayer (red)
  taken at $H = 4$ T and 10 K.  The bilayer TFY signal has
  been scaled by 2 for visibility. (c) The bilayer
  hysteresis derived from the Sm M-edge XMCD in the TFY
  mode, with the field normal geometry. The shaded circles
  show data points while the black line is a smoothed
  average as a guide to the eye. The arrows represent the
  spin-moment (black arrows) and orbital moment (red arrows)
  through the SmN in the bilayer. }
\end{figure*}

We access the magnetic alignment of SmN more directly by turning to
the soft X-ray M$_{4,5}$ edges, which represent $3d\rightarrow 4f$
transitions and thus signal the spin and orbital alignment in the 4$f$
shell. Figure \ref{fig3}(a) sketches the geometry of the M-edge
measurements, with the magnetic field and X-rays parallel to the
surface normal.

Figure \ref{fig3}(b) shows the Sm M$_{4,5}$-edges XMCD in
both TEY and TFY modes for the two samples (see
Supplementary Material for XAS spectra). For the trilayer,
the TFY and TEY spectra in the SmN layer are in excellent
agreement, establishing common SmN 4$f$ alignment in the
near-surface region (TEY) and the bulk (TFY); clearly the
SmN is effectively decoupled from GdN by the LaN blocking
layer. In contrast, both TFY and TEY signals are
substantially weaker in the bilayer, and the TFY signal is
even inverted. The latter is a signature of SmN that is
strongly coupled to the GdN by exchange across the GdN/SmN
interface.

To quantify the differences in XMCD between samples we have
fit the Sm M$_{4,5}$ spectra in the bilayer to that in the
uniformly aligned trilayer. The procedure is justified by
the strong spin-orbit coupling of the 4$f$ electrons, which
keeps spin and orbital moments firmly aligned relative to
each other.\cite{vanderLaan_PRB_1996,Dhesi_PRB_2010} The
XMCD sum rules\cite{Thole_PRL_1992,Carra_PRL_1993} then
imply that the XMCD spectral shape should remain the same
between the samples, with a scaling factor as a measure of
the depth averaged (TFY) and near-surface (TEY) alignment.

Fitting of the spectra yields spin/orbital-alignment ratios
of bilayer-to-trilayer of $R_{\text{TEY}} = 0.20 \pm 0.07$
and $R_{\text{TFY}} = -0.12 \pm 0.02$. For the bilayer then,
the alignment in the surface $\sim 2$ nm probed by TEY is
Zeeman-dominated, though its alignment with the field is
only 20\% of that in bulk SmN. In contrast the average
through the film is of opposite sign, determined by exchange
across the GdN/SmN interface, as was found also in the very
thin SmN layers in the superlattice in the L$_{2,3}$-edge
study above. Clearly there is an inhomogeeous alignment in
the bilayer, a rotation of the spin and orbital moments as
sketched in Figure \ref{fig4}(a).

Figure \ref{fig3}(c) shows an unusual hysteresis curve extracted from
the bulk sensitive TFY measurement of the bilayer, where SmN is
deposited directly on GdN. The same fitting procedure mentioned above
was used to extract the hysteresis. The Sm 4$f$ alignment in this case
shows the same sign inversion seen in the L$_{2,3}$-edge data in
Fig. \ref{fig2}(b), but with diminishing alignment with increasing
fields larger than $\sim 1.5$ T. It is important to notice that in the
field-normal configuration, the shape anisotropy of GdN prevents a
saturated magnetization in applied fields smaller than $\sim 2$ T. Its
magnetization rises approximately linearly with weaker applied fields,
but for larger fields the GdN is saturated; between 2 and 4 T the 4$f$
spins are fully aligned and exert the full Gd-Sm exchange on the SmN
4$f$ spin-moment at the interface. In this region the increasing field
has the effect of modifying the exchange-Zeeman competition which in
turn reduces the bulk averaged XMCD signal as the 4$f$ spin and
orbital moments rotate through the film. In the following section we
pursue deeper insight into the nature of the twisting, or rotating
magnetization.

\section{Analysis and Discussion}
\begin{figure}
\center \includegraphics[width=0.5\columnwidth]{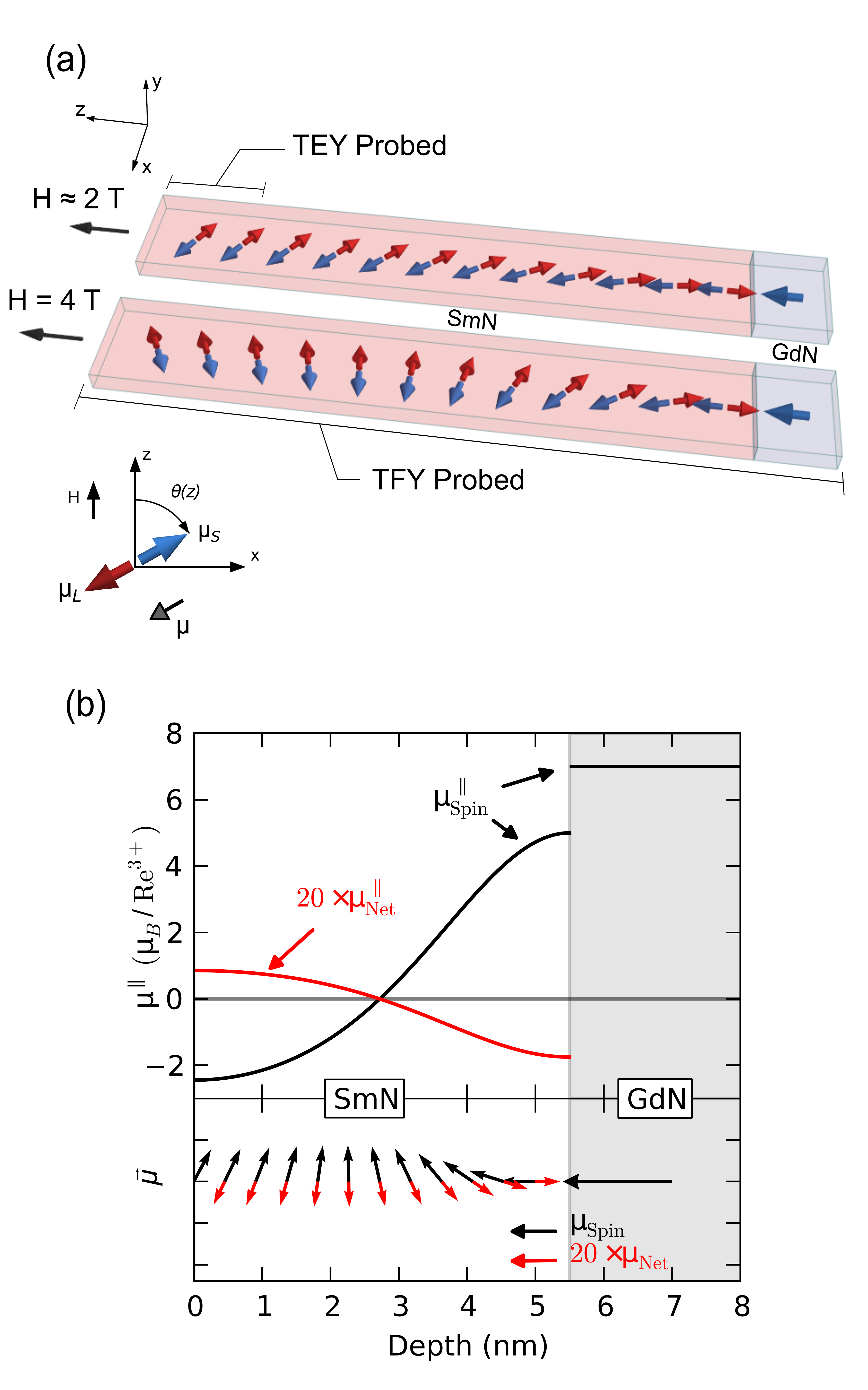}
\caption{\label{fig4} (a) A sketch of the in-plane twisted magnetization
  structure near the SmN-GdN interface with spin moments (blue) and
  orbital moments (red). (b) Calculation of the spin and net moment as
  a function of depth in the SmN layer of the bilayer. The Gd
  moment is fixed at 7 $\mu_B$ through the GdN film, and the Sm
  spin-moment is pinned at the SmN/GdN interface to its maximum value
  of 5 $\mu_B$.}
\end{figure}

In this section we relate the measured TEY and TFY XMCD
results in the bilayer to a model of the twisting SmN
magnetization. We consider a one dimensional model of the
SmN magnetization in the bilayer, in which the resulting
magnetization profile is determined by the balance among (i)
the Sm-Sm exchange energy acting on Sm spin moments, (ii)
the Zeeman energy acting on the SmN net moment, and (iii)
the demagnetization field of SmN.  We note that the shape
anisotropy for SmN is only 0.01 T; under the large fields of
interest here the demagnetization field responsible for the
shape anisotropy can be neglected in comparison to the
Zeeman energy. While anisotropy should play some role, there
are no studies of its effects in SmN, \cite{Meyer_PRB_2008}
and our results suggest it is only a weak
correction. Treating the exchange as acting between atomic
planes parallel to the interface, the total energy per unit
area in a continuum
approximation\cite{Goto_JAP_1965,Thiaville_JMMM_1992} is
then
\begin{align}\label{eq1}
  \mathcal{E} = \int ^d_0 \mathrm{d}z\;\left[A \left(\frac{d\theta
        (z)}{dz} \right) ^2 - \bm{M}_S\cdot\bm{H}\right],
\end{align}
where $A$ is the exchange stiffness, $\bm{M}_S$ is the saturation
magnetization of the SmN, $\bm{H}=H\hat{\bm{z}}$ is the applied field,
and $d=5.5$ nm is the thickness of the SmN film.  $\theta(z)$ is the
depth-varying angle between $\bm{H}$ and the spin-moment $\pmb{\mu}_S$
(see Fig. \ref{fig4}(a)). The Zeeman term adopts the opposite sign as
found in conventional spin-dominant systems because the net moment is
antiparallel to the $\pmb{\mu}_S$, hence $-\bm{M}_S\cdot\bm{H} =
M_SH\cos\theta(z)$. The exchange stiffness $A$ is estimated from the
experimental Curie temperature of SmN using the mean-field
approximation.\cite{Goodenough_1963}

Equation (1) can be minimized analytically to yield the most
energetically favorable configuration, as carried out by
Goto \emph{et al.} \cite{Goto_JAP_1965} for an
exchange-spring system, yielding an expression for
$\theta(z)$ in terms of Jacobi elliptic functions (see
Supplemental Material for details).  The boundary conditions
were chosen such that $\theta(d)=0$ (Sm spin is aligned with
the Gd spin at the SmN-GdN interface) and
$d\theta(z)/dz|_{z=0}=0$ (SmN free surface). These boundary
conditions account for the magnetically soft GdN being
rigidly fixed parallel to the applied field due its large
Zeeman interaction. This fixed GdN then acts as the rigid
pinning layer for the SmN at the interface. We emphasize
that this is in strong contrast to conventional
spin-dominant exchange-spring systems, where the pinning
layer must have a large coercive field in order to remain
rigid because the field is applied \emph{antiparallel} to
its magnetization.

Within the model a twisted phase develops on a scale of
$\ell = \pi/2 \sqrt{2A/HM_S}$; below this thickness a
uniform magnetization ($\theta(z)\equiv 0$) is favored for
given parameters. In an applied field of 4 T this
corresponds to $ \ell \approx 4$ nm, on the order of the SmN
film thickness. Figure \ref{fig4}(b) shows the calculation
of the net moment and spin-moment projected on the $z$-axis
(i.e., $\mu\cos\theta(z)$ and $\mu_S \cos\theta(z)$), as a
function of the depth $z$ in the 4 T field. The length scale
of the twist increases in a field of 2 T as the Zeeman
energy weakens, illustrated in Fig.~\ref{fig4}(a) and the
inset of Fig.~\ref{fig3}(c).

The resulting depth profile of the net SmN moment projected along the
$z$-axis, $\mu \cos \theta (z)$, can be compared to the XMCD spectra
by accounting for the depth-averaging of the XMCD measurement, in
combination with the effective sampling depth in the TXY (TEY or TFY)
measurement schemes, $\lambda_{\text{TXY}}$. The finite sampling depth
$\lambda_{\text{TXY}}$ in the TXY mode results in a detection
efficiency $w_{\text{TXY}} =e^{-z/\lambda_{\text{TXY}}}$ from a depth
$z$. \cite{Stohr_Magnetism_2007} Thus we can approximate the depth
averaged XMCD measurement as returning an effective net moment of
\begin{align}
  \langle \mu\rangle_{\text{TXY}} = \frac{1}{d}\int_0^d \text{d}z \;
  \mu\cos\theta (z)e^{-z/\lambda_{\text{TXY}}}. 
\end{align}

Absolute values of $\mu_S$ and the orbital moment $\mu_L$
can in principle be extracted by applying the XMCD sum
rules, however they require much greater signal-to-noise
ratios than available with the present data. Instead we note
that $\mu_S$ and $\mu_L$ in both samples are fixed
antiparallel by the strong spin-orbit coupling, and the
energy dependence of the XMCD spectra remains unchanged. The
ratios of $\langle \mu\rangle_{\text{TXY}}$ between the
bilayer and trilayer are thus expressions for the
experimentally determined ratios, which simply reflects the
depth-average of $\cos\theta(z)$ within the bilayer. The
calculated ratio $R_{\text{TXY}}= \langle
\mu\rangle_{\text{TXY}} ^{\text{bilayer}}/\langle
\mu\rangle_{\text{TXY}} ^{\text{trilayer}}$ in the TFY mode
gives $R_{\text{TFY}} = -0.11$, in excellent agreement with
the measured ratio of $-0.12 \pm 0.02$. $R_{\text{TFY}}$ is
insensitive to precise value of $\lambda_{\text{TXY}}$
($\approx$ 100 nm), as $d\ll\lambda_{\text{TFY}}$; the
flouresence is effectively unattenuated. The TEY-channel
ratio depends more strongly on $\lambda_{\text{TEY}}$; the
experimental value of $R_{\text{TEY}}=0.20$ is returned for
an electron escape depth of $\lambda_{\text{TEY}} = 2.15$
nm, which is consistent with the expected
range.\cite{Stohr_Magnetism_2007,Thole_PRB_1985} The
excellent agreement of this analytical model with the
experimentally measured spectra thus strongly supports the
mechanism of exchange-Zeeman competition driving the twisted
magnetization in the SmN layer coupled to GdN.

The agreement achieved by using only experimental parameters
and reasonable values of $\lambda_{TXY}$ is encouraging and
indicates that other effects, including bulk and surface
anisotropies are only weak corrections to the exchange and
Zeeman dominated contributions. We add that the continuum
approximation leading to Eq.~\eqref{eq1} has been shown to
be in good agreement with more exact treatments using a
discretized version of the model, even down to a few
monolayers. \cite{Bowden_JPCM_2000}

\section{Conclusion}
In summary, we have observed a novel twisted magnetization phase in a
SmN/GdN bilayer by exploiting the depth dependence of the
electron-yield and flourescence-yield detection modes at the
rare-earth M-edge XMCD. The interfacial pinning of the SmN moment to
GdN was clearly demonstrated in the L-edge XMCD measurements, showing
that the ferromagnetic GdN-SmN exchange coupling is responsible for
the pinning. The decoupling of the SmN and GdN magnetization in the
SmN/LaN/GdN structure points towards magnetic tunnel junctions,
especially attractive within the RENs owing to their epitaxial
compatibility across the series. The appearance of a twisted phase in
the SmN/GdN system also holds intriguing possibilities for spintronics
applications, owing to the semiconducting nature of the pair coupled
with the orbital-dominant magnetism of SmN. For example, the tuning of
the twisted phase length scale $\ell \sim\sqrt{A/HM_S}$ for given
fields can be achieved through doping, or replacement, with other
rare-earth elements, thus modifying the exchange $A$ and the
saturation magnetization $M_S$. \cite{Adachi_Nature_1999} The ability
to control the scale of what is effectively a domain-wall width in
intrinsic ferromagnetic semiconductor heterostructures also allows for
the opportunity to explore spin-orbit torques across controllable
domain-wall widths.

\begin{acknowledgments}
  We acknowledge financial support from the NZ FRST(Grant
  No.~VICX0808) and the Marsden Fund (Grant No.~08-VUW-030). The
  MacDiarmid Institute is supported by the New Zealand Centres of
  Research Excellence Fund. E.A.~thanks the Alexander-von-Humboldt
  foundation for support through a fellowship.
\end{acknowledgments}

\bibliography{library_all.bib}

\end{document}